\documentclass[
aps,%
12pt,%
final,%
notitlepage,%
oneside,%
onecolumn,%
nobibnotes,%
nofootinbib,%
superscriptaddress,%
noshowpacs]%
{revtex4}

\usepackage[T1]{fontenc}
\usepackage[utf8]{inputenc}
\usepackage[english]{babel}
\usepackage{graphicx}
\usepackage{color}
\usepackage{amsmath,amssymb}
\usepackage{bm,mathtools}
\newcommand{\dd}{\mathrm{d}}
\newcommand{\ii}{\mathrm{i}}
\newcommand{\Lag}{\mathcal L}
\newcommand{\Ham}{\mathcal H}
\newcommand{\vev}[1]{\langle #1\rangle}
\newcommand{\ket}[1]{\lvert #1\rangle}
\newcommand{\bra}[1]{\langle #1\rvert}
\newcommand{\uv}{\mathrm{UV}}
\newcommand{\ir}{\mathrm{IR}}

\begin{document}

\noindent {\it Astronomy Reports, 2026, Vol. , No. }
\bigskip\bigskip  \hrule\smallskip\hrule
\vspace{35mm}

\title{POSSIBLE MICROSCOPIC SOURCE OF TORSION\footnote{Paper presented at the Sixth Zeldovich meeting, an international conference in honor of Ya. B. Zeldovich held in Pescara, Italy on July 13--17, 2026. Published by the recommendation of the special editors: R. Ruffini and G. V. Vereshchagin.}}

\author{\bf \copyright $\:$ 2026.\quad \firstname{A.}~\surname{Capolupo,}}%
\email{capolupo@sa.infn.it}
\affiliation{Dipartimento di Fisica ``E.R. Caianiello'', Universit\`a di Salerno, and INFN -- Gruppo Collegato di Salerno, Via Giovanni Paolo II 132, 84084 Fisciano (SA), Italy}%

\author{\bf \firstname{S.}~\surname{Monda,}}%
\email{smonda@unisa.it}
\affiliation{Dipartimento di Fisica ``E.R. Caianiello'', Universit\`a di Salerno, and INFN -- Gruppo Collegato di Salerno, Via Giovanni Paolo II 132, 84084 Fisciano (SA), Italy}%

\author{\bf \firstname{G.}~\surname{Pisacane,}}%
\email{gpisacane@unisa.it}
\affiliation{Dipartimento di Fisica ``E.R. Caianiello'', Universit\`a di Salerno, and INFN -- Gruppo Collegato di Salerno, Via Giovanni Paolo II 132, 84084 Fisciano (SA), Italy}%

\author{\bf \firstname{A.}~\surname{Quaranta,}}%
\email{anquaranta@unisa.it}
\affiliation{University of Camerino, Via Madonna delle Carceri, 62032 Camerino, Italy}%

\author{\bf \firstname{R.}~\surname{Serao}}%
\email{rserao@phy.hr}
\affiliation{Dipartimento di Fisica, Facoltà di Scienze, Università di Zagabria Bijeni\v{c}ka cesta 32, 10000 Zagabria, Croazia}

\begin{abstract}
\centerline{\footnotesize Received: ;$;$ Revised: ;$;$ Accepted: .}\bigskip\bigskip\bigskip Axial fermionic interactions connect vacuum condensation with torsionful backgrounds. We consider a quartic spin spin interaction whose mean field reduction yields a quadratic fermionic Hamiltonian. Its diagonalisation defines a particle and antiparticle condensate, while self consistency determines whether the vacuum develops a non zero axial current. Cutoff, fluctuation and energy conditions delimit the validity of the approximation. When torsion couples to spin, this axial expectation value can provide a microscopic source for an axial torsion background. We then consider mixed Dirac fields on such a background. In the explicit neutrino analysis, metric curvature is neglected while torsion modifies the flavour Bogoliubov coefficients, producing spin dependent oscillation frequencies, condensate densities and CP asymmetries.
\end{abstract}

\maketitle

\section{Introduction}\label{sec:intro}
The relation between fermionic matter and spacetime geometry involves both the coupling of spin to gravity and the quantum structure of the vacuum. Spinors and connections provide the geometrical framework for this relation~\cite{Wald:1984rg,Lawson:1998yr,Penrose:1986ca,Nakahara:2003nw,Baez:1995sj}. General relativity admits formulations in which curvature, torsion and non-metricity play different roles~\cite{BeltranJimenez:2019esp,Aldrovandi:2013wha,Maluf2013,BeltranJimenez:2017tkd,Nester:1998mp,Hayashi:1979qx,Pereira2014,Adak:2005cd,Koivisto:2005yk}, while metric-affine and modified theories extend this geometrical structure and lead to distinct field equations and cosmological applications~\cite{Hehl:1994ue,Hehl:1999sb,Iosifidis:2018zwo,Iosifidis:2021tvx,Iosifidis:2018zjj,Bahamonde:2021gfp,Heisenberg:2018vsk,Jarv:2018bgs}. These classical approaches differ from programmes aimed at quantising gravity~\cite{Rovelli:2004tv,Percacci:2017fkn,tHooft:1993dmi}, including canonical formulations and thermodynamic derivations of gravitational dynamics~\cite{DeWitt:1979er,Jacobson:1995ab}. For a metric-compatible connection, torsion is associated with its antisymmetric part. The coupling of intrinsic spin to an independent connection is central to Einstein-Cartan and Poincaré gauge gravity~\cite{Kibble:1961ba,Sciama:1964wt,Hehl:1976kj,Trautman:2006fp,Blagojevic:2002du,Blagojevic:2013xpa,Obukhov:2006gea,Hehl:2013qga}. Its irreducible components, coupling to Dirac fields and extensions to more general matter and gravitational actions have been extensively studied~\cite{Shapiro2002,Hammond:2002rm,McCrea:1992wa,Arcos:2004tzt,Fabbri:2006xq,Fabbri:2014dxa,Fabbri:2016laz,Mercuri:2006um,Vignolo:2014wva}, together with recent analyses of quantised Dirac fields in torsionful gravity and their cosmological implications~\cite{doi:10.1142/S021988782650012X}. When torsion is algebraic, eliminating it can generate an effective axial four-fermion interaction, providing a possible geometrical origin for the interaction considered below. Spin-dependent gravitational and inertial effects have also been studied through test-particle dynamics, Dirac Hamiltonians, interferometry and rotating or topologically non-trivial backgrounds~\cite{Papapetrou:1951pa,Khriplovich:1997ni,Hehl:1990nf,Obukhov:2000ih,PhysRevD.88.084014,PhysRevLett.61.2639,Bakke:2013sla,Bezerra:2006if}. Constraints on torsion and its relation to effective geometries, defects and non-trivial backgrounds provide complementary perspectives~\cite{Mao:2006bb,Kostelecky:2007kx,Mavromatos:2023wkk,Katanaev:1992kh,deVega:1976xbp,Bezerra:1992rh}. Cosmological studies have considered spin-driven dynamics, singularity avoidance, matter-antimatter asymmetry, modified-gravity phenomenology and possible dark-sector effects~\cite{Trautman:1973wy,Poplawski:2011jz,Poplawski:2010kb,Gasperini:1986mv,Poplawski:2011xf,Tilquin:2011bu,Minkevich:2006ks,Mavromatos:2021hai,Poplawski:2010xfg,Poplawski:2012bw,Mukhanov:2005sc,DeMartino:2017eibi,DeMartino:2017axionicPRL,Capolupo:2026halos}. These results motivate the study of axial sources without selecting a unique cosmological realisation. At the quantum level, the vacuum of an interacting theory need not coincide with the empty state defined by a free-particle basis~\cite{Weinberg:1995mt,Weinberg:1996kr,Srednicki:2007qs,Zee:2003mt,Coleman:2018mew,Zinn-Justin:2002ecy}. Pairing and symmetry breaking provide familiar examples in superconductivity and superfluidity~\cite{Cooper:1956zz,Bardeen:1957mv,Bardeen:1957kj,Allen1983TheoryOS,Bogolyubov:1947zz,ANDERSON:1966bps,Schmitt:2014eka}, while fermion condensates and four-fermion interactions play an important role in strong-interaction physics and have also been studied in curved or bounded backgrounds~\cite{Nambu:1961tp,Klevansky:1992qe,Ioffe:2002ee,Inagaki:1997kz,Ebert:2008zza,Flachi:2013zaa,Bellucci:2011zr}. These systems motivate the use of mean-field methods and Bogoliubov transformations to determine the vacuum structure self-consistently. In QFT on curved spacetime, geometry enters the mode decomposition used to define particles and vacuum states~\cite{Parker:2009uva,DeWitt:1975ys,Ashtekar:1975zn,Parker1969,Parker:2012at,Fulling1973}. Particle creation, acceleration and black-hole radiation are standard examples~\cite{Hawking:1975vcx,Unruh:1976db,Takagi:1986kn,Vanzella:2001ec,Wald1975,Hartle:1976tp,PhysRevD.14.332,Zeldovich:1971mw,Capolupo:2013hawkingUnruh}, while vacuum polarisation and stress tensors require specific regularisation and renormalisation prescriptions~\cite{Christensen:1976vb,Candelas1980,Allen:1985ux,Dowker:1975tf,Hollands:2002ux,Anderson:2007eu,Elizalde:1994gf}. Background electric fields, boundaries and non-trivial topology provide further examples of vacuum modification and induced observables~\cite{Schwinger:1951nm,Nikishov:1970br,Kim:2003qp,Ruffini:2009hg,Plunien:1986ca,Bordag:2009zz,Saharian:2023uxk,Acquaviva:2020cjx}. Rotating geometries offer another setting for spin-dependent fermion dynamics. Frame dragging, rotating black holes and their perturbations provide the classical background~\cite{LenseThirring1918,Kerr1963,Bardeen1972,ChandrasekharBook,Teukolsky:1973ha,Chandrasekhar:1976zz}, while Dirac fields on these geometries have been studied in connection with mode separation, quantisation and fermionic superradiance~\cite{Chandrasekhar:1976ap,Page:1976jj,PhysRevLett.31.1265,10.1063/1.529963,Finster:2000jz,Unruh:1974bw,Casals:2012es,Iyer:1978du,Dai:2023zcj}. Vacuum instability, stress tensors and rotating material sources have also been investigated~\cite{PhysRevD.47.1557,PhysRevD.39.2125,Cardoso:2015zqa}. These effects must be distinguished from those produced by an axial torsion background. Axial currents also appear in quantum anomalies and transport phenomena. Their relation to gauge and gravitational backgrounds is established by the anomaly literature~\cite{Adler:1969gk,BellJackiw1969,Kimura1969,Fujikawa:1979ay,Alvarez-Gaume:1983ihn,Witten:1985xe}, while chiral magnetic and vortical effects, anomalous hydrodynamics and rotating equilibrium states provide related applications~\cite{Vilenkin1979,Vilenkin1980,PhysRevD.78.074033,PhysRevLett.103.191601,PhysRevLett.107.021601,Flachi:2017vlp,Hidaka:2018mel,Buzzegoli:2018wpy,Yee:2014dxa,Stone:2018zel,PhysRevD.92.071501,Prokhorov2018}. In the present problem, however, the axial expectation value arises from a self-consistency condition between a mean field and the vacuum of the corresponding quadratic Hamiltonian. Neutrino mixing introduces a second Bogoliubov structure. In the QFT formulation, the transformation between mass and flavour fields contains particle-antiparticle coefficients and defines a flavour vacuum~\cite{Blasone:2002jv,Capolupo:2004pt}. Extensions to curved backgrounds and bosonic systems show how geometry modifies the corresponding mode overlaps and flavour observables~\cite{Capolupo:2020curvedNeutrinos,Capolupo:2022bosonCurved,Capolupo:2023fieldMixingDM}. The energy content of mixing condensates and possible dark-sector applications have also been studied~\cite{Capolupo:2016pbg,Capolupo:2006et,Capolupo:2019xyd,Capolupo:2021cnt,Capolupo:2023fao,CAPOLUPO2025101894,Capolupo:2025darkUniverse}. The flavour condensate must nevertheless be distinguished from the condensate generated by the self-interaction of a single Dirac field. The analysis below proceeds in two steps. First, an axial four-fermion interaction is treated in the mean-field approximation and the resulting axial current is determined self-consistently~\cite{Capolupo:2024zvm}. Second, three mixed Dirac neutrino fields are quantised on a prescribed torsion background~\cite{Capolupo:2023igw}. When torsion couples to spin, the current generated in the first calculation provides a possible microscopic source for the background used in the second. The general curved-spacetime formulation must be distinguished from the approximation of negligible metric curvature adopted for the explicit neutrino solutions.

\section{Fermion condensates induced by axial self-interaction}\label{sec:self}
The geometrical starting point is a metric-compatible connection whose antisymmetric part is not constrained to vanish. In the notation adopted there, torsion and contorsion are introduced through $\widetilde\Gamma^{\rho}_{\mu\nu}-\widetilde\Gamma^{\rho}_{\nu\mu}=T^{\rho}_{\mu\nu}$ and $\widetilde\Gamma^{\rho}_{\mu\nu}=\Gamma^{\rho}_{\mu\nu}+K^{\rho}_{\mu\nu}$, where $\Gamma^{\rho}_{\mu\nu}$ is the Levi-Civita connection. Metric compatibility in the presence of torsion defines Riemann-Cartan geometry. Einstein-Cartan theory couples this geometrical extension to the intrinsic spin of matter~\cite{Hehl:1976kj,Shapiro2002}, with applications to Dirac fields discussed in Refs.~\cite{Fabbri:2006xq,Fabbri:2014dxa,Fabbri:2016laz,Mercuri:2006um,Vignolo:2014wva}. The torsion tensor can be decomposed into a trace vector, an axial vector and a residual tensor. A minimally coupled Dirac field is sensitive to the axial part. After neglecting metric curvature in the propagation problem, the field equation reads:
\begin{equation}\label{eq:diractorsion}
\ii\gamma^\mu\partial_\mu\psi
=m\psi-\frac32 T^\rho\gamma_\rho\gamma^5\psi.
\end{equation}
The axial vector $T^\rho$ in Eq.~\eqref{eq:diractorsion} must be distinguished from the torsion tensor $T^{\rho}_{\mu\nu}$, being defined through its contraction with the Levi-Civita symbol. A quartic fermionic Lagrangian can contain a spin-spin interaction of torsional origin~\cite{Capolupo:2024zvm}, whose axial expectation value can be determined within a mean-field treatment. \subsection{Quartic Lagrangian and mean-field approximation}
Let us consider the Lagrangian:
\begin{equation}\label{eq:quartic}
\Lag=\frac{\ii}{2}
\left(\bar\psi\gamma^\mu\partial_\mu\psi
-\partial_\mu\bar\psi\gamma^\mu\psi\right)
-m\bar\psi\psi
+\lambda\bar\psi\gamma^5\gamma^\mu\psi
\bar\psi\gamma^5\gamma_\mu\psi,
\end{equation}
where $J_5^\mu=\bar\psi\gamma^5\gamma^\mu\psi$ is the axial current. The coupling $\lambda$ depends on the origin of the interaction and the analysis applies to a generic axial contact term. Since the current is built from the same field whose quantum state is determined, Eq.~\eqref{eq:quartic} describes a genuine self-interaction. The mean-field method reduces this quartic problem to a quadratic one. Thus $\lambda$ specifies the interaction but not the axial expectation value of the vacuum, which must be determined dynamically. In the torsional interpretation the interaction can arise by eliminating non-dynamical torsion in an Einstein-Cartan-type theory. The associated interaction Hamiltonian density is:
\begin{equation}\label{eq:spinham}
\Ham_{\mathrm{Spin\text{-}Spin}}
=-\lambda J_5^\mu J^5_\mu
=-\lambda(J_5^0)^2+\lambda\bm J_5\cdot\bm J_5.
\end{equation}
For negative $\lambda$, the spatial interaction favours spin alignment. The question is whether this interaction can reorganise the vacuum so that it carries a non-zero axial expectation value. The mean field is defined as:
\begin{equation}\label{eq:meanvalue}
L^\mu(t)=\bra{0_C(t)}J_5^\mu\ket{0_C(t)},
\end{equation}
where $\ket{0_C(t)}$ is the condensed vacuum. The expectation value is assumed spatially homogeneous while time dependence is initially allowed. Neglecting fluctuations quadratic in $\delta J_5^\mu=J_5^\mu-L^\mu$ gives:
\begin{equation}\label{eq:meanfield}
J_5^\mu J^5_\mu\simeq2J_5^\mu L_\mu(t)-L_\mu(t)L^\mu(t).
\end{equation}
The quartic Hamiltonian then becomes:
\begin{equation}\label{eq:mfham}
\Ham=\bar\psi(-\ii\bm\gamma\cdot\bm\nabla+m)\psi
-2\lambda L_\mu(t)\bar\psi\gamma^5\gamma^\mu\psi
+\lambda L^\mu(t)L_\mu(t).
\end{equation}
The last term is field independent but contributes to the energy. For a trial $L^\mu(t)$, the quadratic Hamiltonian in Eq.~\eqref{eq:mfham} is diagonalised and its vacuum is used to evaluate the current. Self-consistency requires this expectation value to reproduce $L^\mu(t)$ in Eq.~\eqref{eq:meanvalue}, so the condensate is determined rather than imposed. The corresponding field equation is:
\begin{equation}\label{eq:meanfielddirac}
\ii\partial_0\psi=
\left(-\ii\bm\alpha\cdot\bm\nabla+m\gamma^0\right)\psi
+2\lambda L^0(t)\gamma^5\psi
-2\lambda\bm L(t)\cdot\bm\Sigma\psi.
\end{equation}
Here $\alpha^i=\gamma^0\gamma^i$ and $\bm\Sigma$ is the block-diagonal spin matrix. For fixed $L^\mu(t)$, Eq.~\eqref{eq:meanfielddirac} is linear in $\psi$ while the state dependence remains encoded in the self-consistency condition. The explicit diagonalisation is performed for massless fermions, for which the field operators separate into independent pair sectors and simplify the Bogoliubov construction.
\subsection{Hamiltonian diagonalisation and the condensed vacuum}
Expanding the field in particle and antiparticle modes introduces operators $b_{\bm k,s}$ and $d_{\bm k,s}$, where $s=\pm1$ labels helicity. Since the mean-field Hamiltonian mixes annihilation and creation operators, it is diagonalised through:
\begin{equation}\label{eq:bogoliubov}
\begin{aligned}
B_{\bm k,s}(t)&=\cos\Theta_{\bm k,s}(t)b_{\bm k,s}
-e^{\ii\delta_{\bm k,s}(t)}\sin\Theta_{\bm k,s}(t)d^\dagger_{\bm k,-s},\
D^\dagger_{\bm k,s}(t)&=\cos\Theta_{\bm k,s}(t)d^\dagger_{\bm k,s}
+e^{-\ii\delta_{\bm k,s}(t)}\sin\Theta_{\bm k,s}(t)b_{\bm k,-s}.
\end{aligned}
\end{equation}
The angles $\Theta_{\bm k,s}$ and phases $\delta_{\bm k,s}$ are fixed by diagonalisation. Because the Hamiltonian contains particle-antiparticle pairing terms, the reference vacuum does not generally coincide with the vacuum of the elementary excitations. The operators in Eq.~\eqref{eq:bogoliubov} diagonalise the Hamiltonian while preserving the canonical fermionic algebra. The new vacuum satisfies
\begin{equation}\label{eq:vacuumcondition}
B_{\bm k,s}(t)\ket{0_C(t)}=0,
\qquad D_{\bm k,s}(t)\ket{0_C(t)}=0.
\end{equation}
Introducing the generator $\mathcal R(t)$, it is related to the reference vacuum by
\begin{equation}\label{eq:vacuumgenerator}
\ket{0_C(t)}=\mathcal R^{-1}(t)\ket0.
\end{equation}
The transformation produces a condensate of particle-antiparticle pairs. In particular,
\begin{equation}\label{eq:occupation}
\bra{0_C(t)}b^\dagger_{\bm k,r}b_{\bm q,s}\ket{0_C(t)}
=\delta^3(\bm k-\bm q)\delta_{rs}\sin^2\Theta_{\bm k,r}(t).
\end{equation}
Thus the condensed vacuum has non-zero occupation with respect to the reference particle basis. Anomalous particle-antiparticle expectation values also contribute to the axial current and are required for the self-consistency analysis. Equation~\eqref{eq:occupation} measures the pair content of the new vacuum but does not alone determine the axial current, which also depends on the spin structure and anomalous contractions~\cite{Capolupo:2024zvm}. At finite volume the two vacuum are related by the Bogoliubov generator. In the infinite-volume limit their Fock representations become unitarily inequivalent and the axial current is evaluated in the representation selected by the diagonalised Hamiltonian.
\subsection{Self-consistency equations and energy gap}
The temporal component of the self-consistency condition vanishes after summing over helicities, $L^0=0$. For the spatial components, the integrand is expanded to third order in $\lambda$ and angular integrations are performed. Introducing ultraviolet and infrared cutoffs $Q_\uv$ and $Q_\ir$, the resulting equation is:
\begin{equation}
\label{eq:gap}
\bm L\left[1+\frac{2\lambda}{3\pi^2}(Q_{\uv}^2-Q_{\ir}^2)
+\frac{64\lambda^3}{15\pi^2}\bm L^2
\ln\!\left(\frac{Q_{\uv}}{Q_{\ir}}\right)\right]=0.
\end{equation}
Besides the trivial solution, the non-zero branch satisfies
\begin{equation}\label{eq:gapsolution}
\bm L^2(t)=
-\frac{1+\dfrac{2\lambda}{3\pi^2}(Q_\uv^2-Q_\ir^2)}
{\dfrac{64\lambda^3}{15\pi^2}\ln(Q_\uv/Q_\ir)},
\end{equation}
with coupling range
\begin{equation}\label{eq:couplingrange}
0>\lambda\geq-\frac{3\pi^2}{2(Q_\uv^2-Q_\ir^2)}.
\end{equation}
The negative coupling agrees with the spin-alignment interpretation of Eq.~\eqref{eq:spinham}. Since Eq.~\eqref{eq:gap} is algebraic, it determines the magnitude of the spatial expectation value but not its time evolution. The validity of the mean-field approximation is tested by requiring
\begin{equation}\label{eq:ginzburg}
\left|\vev{J_5^\mu J^5_\mu}-L^\mu L_\mu\right|
\ll\left|L^\mu L_\mu\right|.
\end{equation}
To the order considered and neglecting the infrared cutoff relative to the ultraviolet one, this gives
\begin{equation}\label{eq:cutoff}
Q_\uv^2\simeq\frac{3\pi^2}{2|\lambda|}.
\end{equation}
Thus the gap equation identifies the non-zero branch while the fluctuation criterion restricts the regime in which the mean-field description is reliable. The resulting cutoff lies close to the upper bound obtained from the self-consistency equation. For a gravitational coupling, this scale marks the regime where the low-energy description must be reconsidered. The energetic stability of the condensate is assessed by comparing the same Hamiltonian in the condensed and reference vacuum. The kinetic contribution is
\begin{equation}\label{eq:kineticenergy}
\mathcal E_{\mathrm{KIN}}=
\frac{\lambda^2\bm L^2(t)}{3\pi^2}(Q_\uv^2-Q_\ir^2).
\end{equation}
Including the interaction contribution gives
\begin{equation}\label{eq:energygap}
\Delta E=2V\lambda\bm L^2
\left[1+\frac{\lambda}{6\pi^2}(Q_\uv^2-Q_\ir^2)\right],
\end{equation}
where $V$ is the spatial volume. In the regime specified by Eq.~\eqref{eq:cutoff}, this reduces to
\begin{equation}\label{eq:energyregime}
\Delta E=-\frac32 V|\lambda|\bm L^2.
\end{equation}
Equations~\eqref{eq:energygap} and~\eqref{eq:energyregime} show that the self-consistent condensed vacuum is energetically favoured in the stated regime~\cite{Capolupo:2024zvm}. The axial expectation value selects a preferred spatial direction while the gap equation constrains its magnitude. In a theory where torsion couples to spin density, this expectation value can contribute to the torsion source. The condensed vacuum can therefore provide a quantum source for an axial torsion background. The contact interaction may itself have a torsional origin and the resulting vacuum can modify the source entering the torsion field equation. This interpretation depends on the underlying matter-torsion theory, since a generic four-fermion interaction alone does not define a spacetime geometry.

\section{Quantum fields on torsionful backgrounds and neutrino mixing}\label{sec:mix}
The next step is to study quantum fields in the presence of such a background. The background treatment takes its axial profile as given and determines the fermionic modes and observables on it. Its source may be an axial condensate of the type just discussed, or another spin distribution. This interpretation is consistent with the source-independent setup of the neutrino analysis~\cite{Capolupo:2023igw}.  This separation permits us to explain the possible origin of the background before examining how it affects mixed neutrino fields.
\subsection{Background modes and flavour mixing}
In the framework of QFT in curved spacetime with torsion the Dirac action is given by the following equation:
\begin{equation}\label{eq:curvedaction}
\begin{aligned}
\widetilde S_D={}&\int\dd^4x\sqrt{-g}\left[
\frac{\ii}{2}\left(\bar\psi\gamma^\mu D_\mu\psi
-D_\mu\bar\psi\gamma^\mu\psi\right)-m\bar\psi\psi\right]\\
&+3\int\dd^4x\sqrt{-g}\,T_\mu S^\mu.
\end{aligned}
\end{equation}
Here $D_\mu$ is the torsion-free spin covariant derivative and $S^\mu=\bar\psi\gamma^\mu\gamma^5\psi/2$ is the spin vector used in the source. The first term describes the Dirac field on the metric background; the second contains the axial torsion coupling. Within this approach, quantisation requires a choice of normalised spinor modes solving the background field equation. Their coefficients become particle and antiparticle operators and their inner products determine the transformations between field bases. For mixed fields, these overlaps must be computed between solutions with different masses on the same background. This is why geometry affects the operator construction of flavour mixing, in addition to entering the propagation phases~\cite{Capolupo:2020curvedNeutrinos,Capolupo:2022bosonCurved}.
The explicit neutrino calculation reviewed here uses the further approximation of neglecting metric curvature while retaining axial torsion. The covariant derivatives in Eq.~\eqref{eq:curvedaction} then reduce to ordinary derivatives in the mode problem, giving Eq.~\eqref{eq:diractorsion}. It is therefore an analysis of quantum fields on a torsionful background within the broader curved-spacetime framework, with a simplified metric sector. Curvature is not included in the explicit oscillation curves of the source article. The background used for propagation is prescribed. A vacuum axial current generated by the four-fermion self-interaction offers a possible source motivation, but the neutrino mode functions do not require that source to be fixed in advance. Once the profile is specified, the calculation quantises three Dirac mass fields and constructs their flavour combinations. This change of task is the transition from determining an axial mean field to determining the quantum response of mixed neutrinos to a torsion background.
For the constant background the axial torsion is directed along the third spatial axis. The mode prescription introduces spin-dependent masses and energies as follows:
\begin{equation}\label{eq:publishedmodes}
\widetilde m_i^{\pm}=m_i\pm\frac32 T^3,
\qquad E_{\bm k,i}^{\pm}=\sqrt{\bm k^2+(\widetilde m_i^{\pm})^2},
\qquad N_i^{\pm}=\sqrt{\frac{E_{\bm k,i}^{\pm}+\widetilde m_i^{\pm}}{2E_{\bm k,i}^{\pm}}}.
\end{equation}
We use these quantities to summarise the expressions reported in that work. The labels $+$ and $-$ distinguish the two spin orientations employed in its calculation. The mode normalisation is chosen so that the positive- and negative-frequency spinors have unit norm.
The relation between flavour and mass fields iis given as follow:
\begin{equation}\label{eq:mixingmatrix}
\begin{pmatrix}\nu_e\\\nu_\mu\\\nu_\tau\end{pmatrix}
=\begin{pmatrix}
c_{12}c_{13}&s_{12}c_{13}&s_{13}e^{-\ii\delta}\\
-s_{12}c_{23}-c_{12}s_{23}s_{13}e^{\ii\delta}
&c_{12}c_{23}-s_{12}s_{23}s_{13}e^{\ii\delta}&s_{23}c_{13}\\
s_{12}s_{23}-c_{12}c_{23}s_{13}e^{\ii\delta}
&-c_{12}s_{23}-s_{12}c_{23}s_{13}e^{\ii\delta}&c_{23}c_{13}
\end{pmatrix}
\begin{pmatrix}\nu_1\\\nu_2\\\nu_3\end{pmatrix},
\end{equation}
where $c_{ij}=\cos\theta_{ij}$, $s_{ij}=\sin\theta_{ij}$ and $\delta$ is the Dirac phase. In the operator formulation, the transformation is implemented by the ordered generator $I_\theta(t)=I_{23}(t)I_{13}(t)I_{12}(t)$. The vacuum relation at finite volume is
\begin{equation}\label{eq:flavourvacuum}
\ket{0(t)}_{e,\mu,\tau}=I_\theta^{-1}(t)\ket0_{1,2,3}.
\end{equation}
The mass vacuum in Eq.~\eqref{eq:flavourvacuum} is defined using the particle operators of the background modes. The flavour vacuum is then obtained by applying the mixing generator to that state. These definitions distinguish the role of the geometrical background from the role of the flavour transformation: torsion enters the mode solutions, while the mixing angles specify how mass fields are combined. The generator retains the operator structure of the three-flavour construction in flat spacetime~\cite{Blasone:2002jv}. Here it contains fields solving the torsion-dependent Dirac equation. Consequently, the spinor inner products entering the flavour annihilators differ from the torsion-free ones. The electron-neutrino annihilator is:
\begin{equation}\label{eq:electronoperator}
\begin{aligned}
\alpha^r_{\bm k,\nu_e}(t)={}&c_{12}c_{13}\alpha^r_{\bm k,1}
+s_{12}c_{13}\left[(\Gamma^{rr}_{12;\bm k}(t))^*\alpha^r_{\bm k,2}
+\varepsilon^r\Sigma^{rr}_{12;\bm k}(t)\beta^{r\dagger}_{-\bm k,2}\right]\\
&+e^{-\ii\delta}s_{13}\left[(\Gamma^{rr}_{13;\bm k}(t))^*\alpha^r_{\bm k,3}
+\varepsilon^r\Sigma^{rr}_{13;\bm k}(t)\beta^{r\dagger}_{-\bm k,3}\right].
\end{aligned}
\end{equation}
Here $\varepsilon^\pm=\mp1$ is the spin-dependent sign used in the paper. The coefficients $\Gamma$ and $\Sigma$ express the particle and particle--antiparticle overlaps, respectively. Equation~\eqref{eq:electronoperator} displays the Bogoliubov structure responsible for the flavour-vacuum condensate.

\subsection{Bogoliubov coefficients and flavour oscillations}
In the momentum frame adopted in its calculation, mixed-spin overlaps vanish in the case of costant torsion. The diagonal particle overlap is reported as
\begin{equation}\label{eq:xi}
\Xi^{\pm\pm}_{ij;\bm k}
=N_i^\pm N_j^\pm\left[
1+\frac{k^2}{(E_{\bm k,i}^\pm+\widetilde m_i^\pm)
(E_{\bm k,j}^\pm+\widetilde m_j^\pm)}\right].
\end{equation}
The particle--antiparticle overlap $\chi^{rs}_{ij;\bm k}$ is defined by the inner product of a positive-frequency mode of one mass and a negative-frequency mode of the other. Both coefficients depend on spin through the modes in Eq.~\eqref{eq:publishedmodes}. Their canonical normalisation is expressed in the source by
$\label{eq:canonical}
\sum_r\left(|\Xi^{\pm r}_{ij;\bm k}|^2+|\chi^{\pm r}_{ij;\bm k}|^2\right)=1.$ For stationary modes, their time-dependent phases involve the difference and sum of the corresponding energies:
$\Xi^{rs}_{ij;\bm k}(t)=\Xi^{rs}_{ij;\bm k}
 e^{\ii(E^s_{\bm k,j}-E^r_{\bm k,i})t},$ $
\chi^{rs}_{ij;\bm k}(t)=\chi^{rs}_{ij;\bm k}
 e^{\ii(E^s_{\bm k,j}+E^r_{\bm k,i})t}.$
These two frequency combinations subsequently appear in the oscillation formulae. The contribution associated with $\chi$ is a consequence of field mixing and is absent from the corresponding quantum-mechanical description.

The flavour charges are constructed by applying the mixing generator to the mass charges. In terms of flavour operators, gives:
\begin{equation}\label{eq:flavourcharge}
::Q_{\nu_\sigma}(t)::=
\sum_r\int\dd^3k\left[
\alpha^{r\dagger}_{\bm k,\nu_\sigma}(t)\alpha^r_{\bm k,\nu_\sigma}(t)
-\beta^{r\dagger}_{\bm k,\nu_\sigma}(t)\beta^r_{\bm k,\nu_\sigma}(t)\right],
\qquad\sigma=e,\mu,\tau.
\end{equation}
The double colon denotes normal ordering with respect to the flavour vacuum. Oscillations are obtained by evaluating these charges on a state prepared with a definite flavour at the initial time. For spin up, the anticommutator expression is:
\begin{equation}\label{eq:chargeexpectation}
\mathcal Q^{\bm k}_{\uparrow\nu_\rho\to\nu_\sigma}(t)
=\left|\left\{\alpha^\uparrow_{\bm k,\nu_\sigma}(t),
\alpha^{\uparrow\dagger}_{\bm k,\nu_\rho}(0)\right\}\right|^2
+\left|\left\{\beta^{\uparrow\dagger}_{-\bm k,\nu_\sigma}(t),
\alpha^{\uparrow\dagger}_{\bm k,\nu_\rho}(0)\right\}\right|^2.
\end{equation}
The analogous expression holds for spin down. Their oscillatory terms are governed by
\begin{equation}\label{eq:frequencies}
\Delta^r_{ij;\bm k}=\frac{E^r_{\bm k,j}-E^r_{\bm k,i}}{2},
\qquad\Omega^r_{ij;\bm k}=\frac{E^r_{\bm k,j}+E^r_{\bm k,i}}{2}.
\end{equation}
The squared Bogoliubov coefficients weight the terms containing these frequencies. This is the central result: torsion affects the amplitudes through the mode overlaps, in addition to affecting the frequencies through the energy splitting. Note that the constant-background results show different oscillation curves for the two spin orientations. The analysis identifies low momenta and torsion scales comparable to the neutrino mass and momentum scales as the regime in which the effect is most pronounced. In the torsion-free ultrarelativistic limit, the formulae reduce to the usual Pontecorvo expressions. The source additionally discusses how the spin dependence changes as the torsion scale is increased. In the time-dependent torsion ansatz, the analytical structure of the solutions is similar, but the Bogoliubov modes differ.

\subsection{Spin-dependent flavour condensate and CP asymmetry}
The flavour vacuum is a condensate already in the torsion-free field-theoretic treatment of mixing~\cite{Capolupo:2004pt,Blasone:2002jv}. The new feature investigated here is the dependence of its occupation densities on spin orientation. These densities are obtained by evaluating mass-field number operators on the flavour vacuum. For the two generation fields, the  expressions are given as follow:
\begin{equation}\label{eq:flavourdensities}
\begin{aligned}
N^{r}_{1;\bm k}&=s_{12}^2c_{13}^2|\Sigma^{rr}_{12;\bm k}|^2
+s_{13}^2|\Sigma^{rr}_{13;\bm k}|^2,\\
N^{r}_{2;\bm k}&=\left|-s_{12}c_{23}+e^{\ii\delta}c_{12}s_{23}s_{13}\right|^2
|\Sigma^{rr}_{12;\bm k}|^2
+s_{23}^2c_{13}^2|\Sigma^{rr}_{23;\bm k}|^2.
\end{aligned}
\end{equation}
Here $r$ is either of the two spin orientations. The same density is obtained for particles and antiparticles of each mass field. 
The effect of torsion on $CP$ asymmetry is related to antineutrino states with a negative flavour charge; the definition based on charge therefore involves a sum:
\begin{equation}\label{eq:cpdefinition}
\Delta^{\rho\sigma}_{r;CP}(t)
=\mathcal Q^{\bm k}_{r\nu_\rho\to\nu_\sigma}(t)
+\mathcal Q^{\bm k}_{r\bar\nu_\rho\to\bar\nu_\sigma}(t).
\end{equation}
The intrinsic mixing-phase factor is the Jarlskog invariant, written in the parameterisation used there as:
\begin{equation}\label{eq:jarlskog}
J_{CP}=\frac18\sin\delta\,
\sin(2\theta_{12})\sin(2\theta_{13})\cos\theta_{13}\sin(2\theta_{23}).
\end{equation}
Its Bogoliubov coefficients and oscillation frequencies depend on spin, so the two spin orientations give different asymmetries. The calculations of the flavour condensate, oscillation charges and $CP$ asymmetry consequently describe related consequences of the same torsion-dependent mode overlaps. The result established by the calculation is the spin dependence within the analysed backgrounds, while its observational application is a prospect discussed in that work.

\section{Conclusions}\label{sec:conclusions}
The four-fermion analysis establishes a self-consistent mechanism for generating an axial current in the fermion vacuum~\cite{Capolupo:2024zvm}. The quartic interaction is reduced to a quadratic Hamiltonian through a trial axial mean field. A Bogoliubov transformation constructs the corresponding particle--antiparticle vacuum, and the expectation value of the current in this vacuum must reproduce the assumed mean field. This last condition determines the condensate: the axial current is an outcome of the self-interaction calculation. In the massless, spatially homogeneous approximation, the temporal expectation value vanishes, while a non-trivial spatial solution exists in the negative-coupling range. Its magnitude depends on the momentum cutoffs. The fluctuation estimate specifies the regime supporting the mean-field approximation, within which the interaction energy gain exceeds the kinetic cost of pair formation. These conditions are essential to the interpretation of the lower-energy condensed vacuum. In a theory coupling torsion to spin, the resulting axial current can contribute to the torsion source. The four-fermion mechanism therefore offers one possible justification for a torsional background. The subsequent analysis is QFT in a curved spacetime with torsion: the geometric background fixes the Dirac modes used to quantise the fields. Although the covariant formulation includes metric curvature, the neutrino calculations neglect metric-curvature effects and retain the prescribed torsion. The two studies thus address successive, distinct problems; they do not solve the coupled evolution of the condensate, geometry and mixed fields. The mixing calculation shows how torsion enters both the propagation phases and the spinor overlaps governing the flavour Bogoliubov transformation~\cite{Capolupo:2023igw}. For the constant and linearly time-dependent backgrounds examined, this produces spin-dependent flavour charges, condensate occupation densities and CP asymmetries. The low-momentum regime is the principal setting discussed for the relevance of these field-theoretic effects. The flavour condensate arises from the relation between mass and flavour fields, whereas the axial condensate follows from the self-interaction consistency condition. 
\label{sec:mainend}

\section*{Acknowledgments}
A.C. and S.M. acknowledges partial support from MIUR and INFN. A.C. also acknowledges COST Action CA1511, Cosmology and Astrophysics Network for
Theoretical Advances and Training Actions (CANTATA). R.S. acknowledges support from the project ``Implementation of cutting-edge research and its application as part of the Scientific Center of Excellence for Quantum and Complex Systems, and Representations of Lie Algebras'', Grant No. PK.1.1.10.0004, co-financed by the European Union through the European Regional Development Fund - Competitiveness and Cohesion Programme 2021-2027.

\end{document}